\documentclass[aps,prb,reprint,superscriptaddress,amsmath,amssymb,showpacs]{revtex4-1}
\usepackage{graphicx}

\begin{document}
\title{Effect of hydrogen adsorption on the quasiparticle spectra of graphene}

\author{M. Farjam}
\affiliation{School of Nano-Science, Institute for Research in Fundamental
Sciences (IPM), P.O. Box 19395-5531, Tehran, Iran}

\author{D. Haberer}
\affiliation{IFW Dresden, P.O. Box 270116, D-01171 Dresden, Germany}

\author{A. Gr\"uneis}
\affiliation{IFW Dresden, P.O. Box 270116, D-01171 Dresden, Germany}
\affiliation{Faculty of Physics, University of Vienna, Strudlhofgasse 4,
1090 Vienna, Austria}

\date{\today}

\begin{abstract}

We use the non-interacting tight-binding model to study the effect of isolated
hydrogen adsorbates on the quasiparticle spectra of single-layer graphene.
Using the Green's function approach, we obtain analytic expressions for the
local density of states and the spectral function of hydrogen-doped graphene,
which are also numerically evaluated and plotted.  Our results are relevant
for the interpretation of scanning tunneling microscopy and angle-resolved
photoemission spectroscopy data of functionalized graphene.

\end{abstract}


\maketitle

The adsorption of hydrogen on graphene is of fundamental as well as applied
significance.  Current interest in hydrogenated graphene is due to the
possibility of opening a band gap in the semimetallic graphene, which is
essential for applications in electronics. This has recently prompted several
research groups to experiment with hydrogenated graphene and to study its
electronic and transport properties.  \cite{elias2009, bostwick2009,
balog2010, haberer2010} Calculations based on density-functional theory (DFT)
gave the motivation to search for hydrogenated derivatives of graphene,
since they showed that \emph{graphane}, i.e., the fully hydrogenated compound,
is both a stable 2D hydrocarbon and a band insulator.  \cite{sofo2007} Earlier
DFT calculations had studied the effect of hydrogen adsorption on the
electronic structure of graphene, using a $4\times4$ regular arrangement,
i.e., an H:C ratio of 1:32, and found a band gap opening and the appearance of
a spin-polarized dispersionless midgap band.  \cite{duplock2004}

In this work, we analyze the effect of a single hydrogen impurity on the
electronic structure of graphene. As we show below, the analysis can be
extended also to calculate some properties of randomly distributed H
adsorbates at a low-density coverage.  A hydrogen atom adsorbed on graphene
forms a covalent bond atop a carbon atom, and changes the orbital state of the
host atom from $sp^2$ to $sp^3$.  This is expected to have a characteristic
effect on the local density of states (LDOS), and to produce Friedel
oscillations in the surrounding charge density. These properties can be mapped
out by scanning tunneling microscopy (STM) and, indeed, hydrogen monomers and
dimers on graphene have been imaged by this technique.  \cite{hornekaer2006,
guisinger2009} As the graphene sample is exposed to a hydrogen plasma, we
initially expect the adsorption of a low concentration of randomly distributed
H atoms, and in later stages more complicated and perhaps regular adsorption
arrangements. We can enumerate several changes that can be expected in the
initial stages of adsorption to occur in the quasiparticle spectra.  First,
due to scattering from the H impurities there is an effect of line broadening
in the momentum-resolved spectral function.  Second, H adsorption breaks the
sublattice symmetry which can lead to a band gap opening at the Dirac point.
Third, since hydrogen impurity on graphene acts as a resonant scatterer, a
midgap resonance appears in the density of states.  \cite{pereira2006} The
spectral function is accessible by angle-resolved photoemission spectroscopy
(ARPES), and the above changes have already been obeserved in experiments on
hydrogenated graphene.  \cite{bostwick2009, balog2010, haberer2010}

Theoretical studies of the electronic and transport properties of hydrogenated
graphene have been based on a tight-binding model which describes the
$\pi$ bands of graphene and an additional orbital for the H adsorbate.
\cite{robinson2008, wehling2010, yuan2010} The purpose of this Brief Report is
to calculate the local density of states and the spectral function of sparsely
hydrogenated graphene within the framework of this tight-binding model.  As in
earlier theoretical studies of the effect of impurities on the spectral
properties of graphene, a Green's function analysis can be used to obtain the
desired quantities.  \cite{bena2005, *bena2008, pereira2006, pereira2008,
skrypnyk2006, *skrypnyk2007, wehling2007, feher2009, bacsi2010, skrypnyk2011}
However, in contrast with most previous works which have considered
substitional impurities, we deal with adsorbate impurities.

We begin our derivation by defining the non-interacting tight-binding model
that describes the $\pi$ bands of graphene with an additional adsorbed
hydrogen atom as \cite{robinson2008, yuan2010}
\begin{equation} \label{H}
\mathcal{H}=\mathcal{H}_0+\mathcal{H}_1+\mathcal{H}_2,
\end{equation}
where $\mathcal{H}_0$ describes pure graphene,
\begin{equation} \label{H0}
\mathcal{H}_0=-t\sum_{\langle{i,j}\rangle} c_i^\dag c_j,
\end{equation}
$\mathcal{H}_1$ describes an isolated hydrogen atom,
\begin{equation} \label{H1}
\mathcal{H}_1=\varepsilon_d d^\dag d,
\end{equation}
and $\mathcal{H}_2$ describes the hybridization between graphene and hydrogen
adsorbed on an arbitrary host site specified by index $l$,
\begin{equation}
\mathcal{H}_2=V (c^\dag_l d + d^\dag c_l).
\end{equation}
Interactions involving spin have not been included in this model, and the spin
degree of freedom has been omitted for simplicity. The energy parameters which
have been obtained from fits to first-principles band structure calculations
are given by $t=2.6$ eV, $V=-2t$ and $\varepsilon_d=-t/16$.
\cite{wehling2010}

\begin{figure}
\includegraphics[width=\columnwidth]{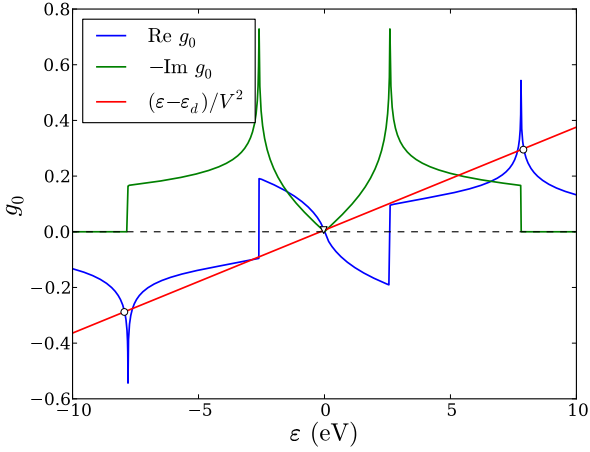}
\caption{\label{fig1} (Color online) Graphical solution of
$\varepsilon-\varepsilon_d-V^2\,\mathrm{Re}\,g_0(\varepsilon)=0$ showing the
positions of the resonance just to the left of Dirac point, and two localized
states just below and above energy bands, respectively.}
\end{figure}

The local density of states at a given site $m$ is given by
\begin{equation} \label{ldos}
\rho(m;\varepsilon)=-\frac{1}{\pi}\mathrm{Im}\,G(m,m;\varepsilon),
\end{equation}
where $G(m,m;\varepsilon)$ is a matrix element of the Green's function
$\mathcal{G}(\varepsilon)=(\varepsilon-\mathcal{H}+i0^+)^{-1}$, while
the total DOS of the system is given by
\begin{equation} \label{dos}
\rho(\varepsilon)=-\frac{1}{\pi}\mathrm{Im}\,\mathrm{Tr}\,\mathcal{G}
(\varepsilon).
\end{equation}
Therefore, we must calculate the Green's function corresponding to
Eq.~(\ref{H}).  The crucial step is to consider the graphene or the hydrogen
atom as a separate entity, and incorporate the presence of the other system in
its Hamiltonian in terms of a self-energy.

The self-energy in a device, is generally given by
$\mathcal{V}{g_R}\mathcal{V}^\dag$
where $\mathcal{V}$ is the coupling Hamiltonian and $g_R$ is the subset of the
Green's function of the reservoir which makes the interface with the device.
\cite{datta2005} Thus instead of Eq.~(\ref{H1}), we can describe a hydgrogen
atom in contact with graphene by
\begin{equation} \label{tH1}
\tilde{\mathcal{H}}_1=\left[\varepsilon_d+V^2 g_0(\varepsilon)\right]
d^\dag d,
\end{equation}
where the additional term is the self-energy induced by graphene and
$g_0(\varepsilon)$ is the diagonal matrix element of the Green's function of
graphene which describes the host site.  On the other hand, for graphene in
contact with hydrogen, the Hamiltonian becomes \cite{robinson2008}
\begin{equation} \label{tH0}
\tilde{\mathcal{H}}_0=-t\sum_{\langle{i,j}\rangle} c_i^\dag c_j
+\frac{V^2}{\varepsilon-\varepsilon_d} c_l^\dag c_l,
\end{equation}
where the additional term is the self-energy induced by hydrogen at the host
site, which has the general form if we note that
$1/(\varepsilon-\varepsilon_d)$
is the Green's function of an isolated hydrogen.  We can now calculate the
LDOS from the Green's functions corresponding to Eqs.~(\ref{tH1}) and
(\ref{tH0}), respectively.  From Eq.~(\ref{tH1}) it follows that the LDOS at
the hydrogen site is given by
\begin{equation} \label{rhoh}
\rho_\mathrm{H}(\varepsilon)=-\frac{1}{\pi}\mathrm{Im}\,\frac{1}
{\varepsilon-\varepsilon_d-V^2 g_0(\varepsilon)}.
\end{equation}

\begin{figure*}
\includegraphics[width=1.4\columnwidth]{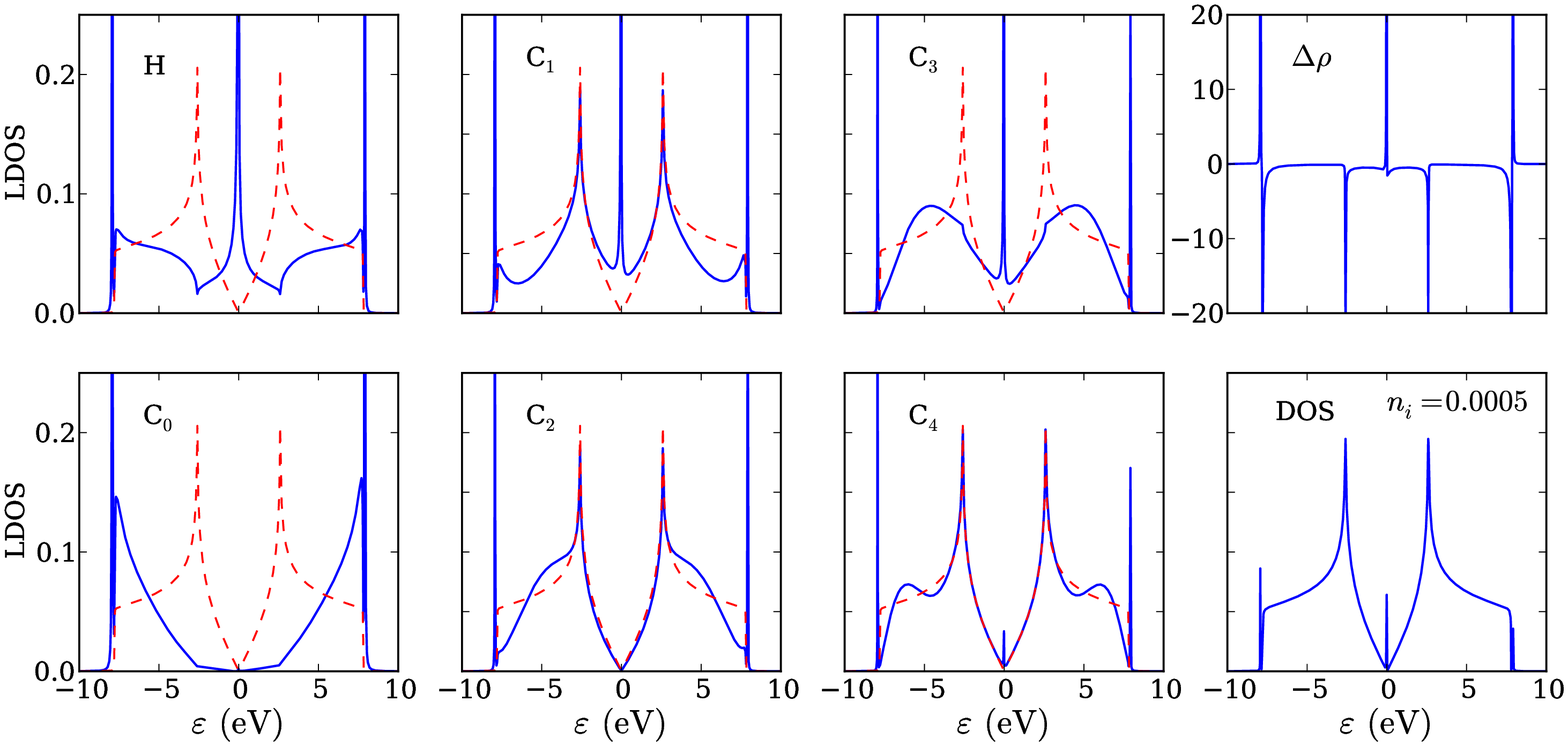}%
\hspace{.1in}
\includegraphics[width=2in]{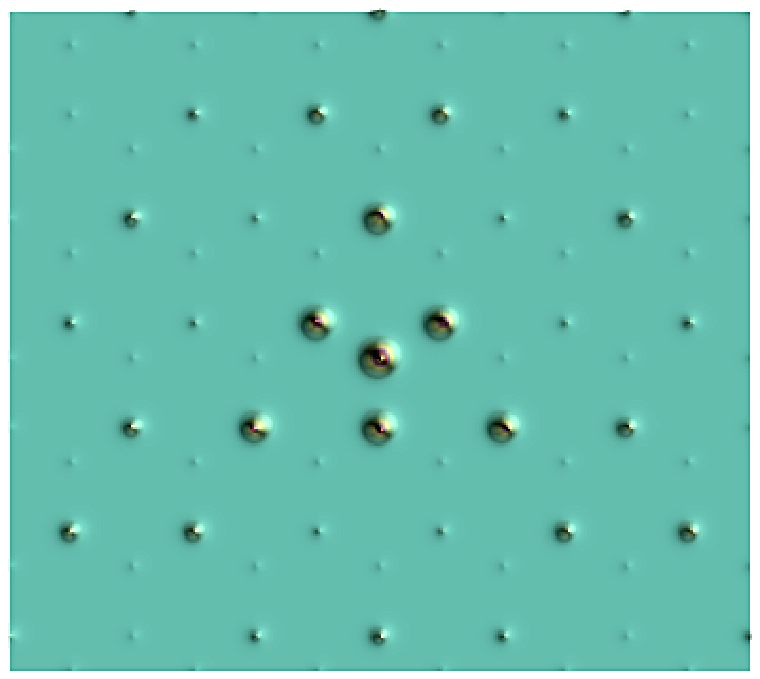}
\caption{\label{fig2} (Color online) Local density of states of graphene with
a single hydrogen adsorbate. (Left panel) The subplots show LDOS at the H site,
the host site, C$_0$, and its four nearest neighbors, denoted by C$_1$--C$_4$,
as well as the change in total DOS of graphene, and the DOS per carbon atom
for 0.05\% hydrogen coverage.  (Right panel) LDOS at the H site (at center) and
nearby graphene sites of a rectangular sample of $6\times6$ unit cells, for an
energy near the resonance $\approx-0.03$\,eV.}
\end{figure*}

Equation (\ref{tH0}) is now similar to the Hamiltonian of a  substitutional
impurity, with an energy-dependent on-site parameter, so the corresponding
Green's function can be obtained exactly from \cite{economou2006}
\begin{equation} \label{tG0}
\tilde{\mathcal{G}}_0=\mathcal{G}_0+\mathcal{G}_0\mathcal{T}\mathcal{G}_0,
\end{equation}
where the $t$-matrix is given by \cite{robinson2008}
\begin{equation} \label{tmat}
\mathcal{T}=\frac{V^2}
{\varepsilon-\varepsilon_d-V^2 g_0(\varepsilon)}\,c^\dag_l{c}_l.
\end{equation}
We can find the positions of the resonances and localized states by setting
the denominators in Eqs.~(\ref{rhoh}) and (\ref{tmat}) equal to zero,
\begin{equation} \label{poles}
\varepsilon-\varepsilon_d-V^2\,\mathrm{Re}\,g_0(\varepsilon)=0.
\end{equation}
A graphical solution of Eq.~(\ref{poles}) is shown in Fig.~\ref{fig1}.  The
position of a solution just to the left of the Dirac point implies a sharp
resonance since $\mathrm{Im}\,g_0$, which measures the width of the resonance,
is vanishingly small.  It is to be noted that the strong hybridization
between H and C, parameterized as $V$, is responsible for the resonance being
much closer to the Dirac point than the on-site energy, $\varepsilon_d$.  The
other solutions of Eq.~(\ref{poles}) correspond to two localized states with
energies just below and above the band limits, respectively.  It is
interesting to compare this picture with the case of a vacancy which is also a
resonant scatterer. For a vacancy, the equation to solve would change to
$1/U=\mathrm{Re}\,g_0(\varepsilon)$ where $U$ would now be a large positive
constant representing the on-site energy of a vacancy.  There will again be
a resonance state near the neutrality point, but only one localized state
outside the bands at a large positive energy.
(cf.~Ref.~\onlinecite{pereira2008})

From Eq.~(\ref{tG0}), it is easy to show that the LDOS at a given site of
graphene is given by
\begin{equation} \label{rhom}
\rho(m;\varepsilon)=-\frac{1}{\pi}\mathrm{Im}\,\frac{(\varepsilon
-\varepsilon_d)g_0+V^2(g_{ml}g_{lm}-g_0^2)}{\varepsilon-\varepsilon_d
-V^2 g_0},
\end{equation}
where $g_{ml}\equiv\langle m|\mathcal{G}_0|l\rangle$, and for the honeycomb
lattice we can use $g_{ml}=g_{lm}$. In particular, for the
host site $m=l$, and with $g_{ll}=g_0$, we find
\begin{equation} \label{rhol}
\rho(l;\varepsilon)=-\frac{1}{\pi}\mathrm{Im}\,\frac{(\varepsilon
-\varepsilon_d) g_0}{\varepsilon-\varepsilon_d-V^2 g_0}.
\end{equation}
When the site $m$ is very far from the host site $l$, $g_{ml}\rightarrow0$, and
$\rho(m;\varepsilon)\rightarrow(-1/\pi)\mathrm{Im}g_0(\varepsilon)=\rho_0
(\varepsilon)$, i.e., it approaches the LDOS of pure graphene.

Evaluation of the LDOS at any lattice site by Eqs.~(\ref{rhom}) and
(\ref{rhol}) requires numerical values of the lattice Green's functions $g_0$
and $g_{lm}$, which can be conveniently calculated by the expressions given by
Horiguchi in terms of elliptic functions. \cite{horiguchi1972}  In
Fig.~\ref{fig2}, we have plotted the LDOS at the host site and its four
nearest neighbors, which show that the resonance is small on the sites of the
sublattice of the host, and large on the other sublattice, which seems to be a
common property that is shared by vacancies also.  \cite{wehling2007}
Figure~\ref{fig2} shows that at sites farther from the impurity the LDOS
approaches that of pure graphene in an oscillatory fashion.

From Eq.~(\ref{tG0}) it follows that the change in the total DOS of graphene
is given by
\begin{equation}
\Delta\rho=\frac{1}{\pi}\mathrm{Im}\left(\frac{dg_0}{d\varepsilon}
\frac{V^2}{\varepsilon-\varepsilon_d-V^2 g_0}\right).
\end{equation}
To calculate the DOS corresponding to Eq.~(\ref{H}), since we need a finite
number of impurities to make the effect finite, we assume to have $N$ carbon
atoms in the graphene, and $N_i\lll{N}$ well isolated hydrogen adsorbates. We
can then write the DOS per carbon site as
\begin{equation} \label{rho}
\rho(\varepsilon)=\rho_0(\varepsilon)+n_i[\Delta\rho(\varepsilon)+
\rho_\mathrm{H}(\varepsilon)],
\end{equation}
where $\rho_0$ is the DOS of pure graphene, and $n_i=N_i/N$ is the ratio of
hydrogen to carbon atoms.  A plot of $\rho_H$ is shown in Fig.~\ref{fig2}
where the expected resonance features can be observed. We have also plotted the
change in the total DOS, $\Delta\rho$, and the DOS for graphene with a tiny
amount of 0.05\% of hydrogen. We note that the total DOS is in agreement with
those obtained by numerical calculations. \cite{yuan2010} It must be remarked
that the values of $\Delta\rho$ can become negative while the total DOS must
always remain positive.  The explanation why this is not unphysical is that
the measurable quantity, which is the total DOS, remains positive, because it is
of $\mathcal{O}(N)$ while the change in it due to the presence of a single H
adsorbate is of $\mathcal{O}(1)$.  Also shown in Fig.~\ref{fig2} is the LDOS
at the sites of a rectangular sample of $6\times6$ unit cells of graphene, at
an energy near the resonance.  This result shows the threefold symmetry and
fluctuations of LDOS, and can be related to STM images.

\begin{figure}
\includegraphics[width=\columnwidth]{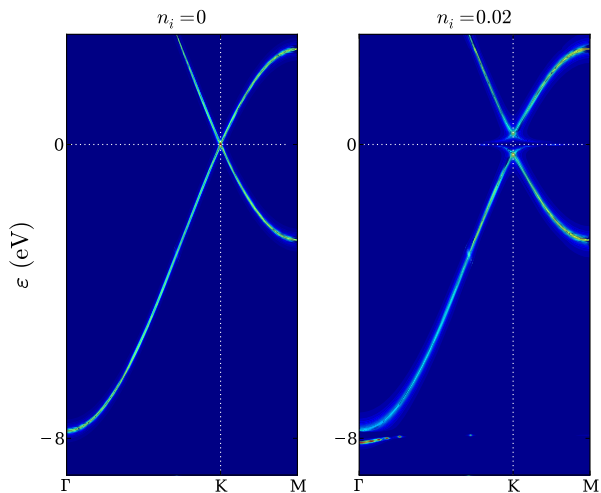}
\caption{\label{fig3} (Color online) Spectral function of graphene along the
$\Gamma$KM path of its Brillouin zone. (Left panel) Clean graphene.
(Right panel)
Graphene with 2\% hydrogen adsorption. General line broadening, appearance of
a localized state just below the band, and changes similar to a gap opening
near zero energy are the noteworthy features.}
\end{figure}

It is easy to extend the above results to find the effect of a low density of
H adsorbates on the spectral function, given by
\begin{equation}
A(\mathbf{k},\varepsilon)=\frac{-2\,
\mathrm{Im}\,\Sigma(\mathbf{k},\varepsilon)}
{[\varepsilon-\varepsilon_\mathbf{k}-\mathrm{Re}\,\Sigma(\mathbf{k},
\varepsilon)]^2+[\mathrm{Im}\,\Sigma(\mathbf{k},\varepsilon)]^2},
\end{equation}
where $\Sigma(\mathbf{k},\varepsilon)$ is the self-energy due to the presence
of hydrogen impurities and $\varepsilon_\mathbf{k}$ is the energy dispersion
of graphene,
\begin{equation}
\varepsilon_\mathbf{k}=\pm t\left[{3+2\left(\cos k_xa+2\cos\frac{k_xa}{2}
\cos\frac{k_ya\sqrt{3}}{2}\right)}\right]^{1/2},
\end{equation}
where $a=2.46$ \AA\ is graphene lattice constant. For low concentrations of H
adsorbates, the average $t$-matrix approximation (ATA) applies and, noting
that here we have no explicit $\mathbf{k}$ dependence, we can write
\begin{equation} \label{ata}
\Sigma(\varepsilon)=\frac{n_iV^2}{\varepsilon-\varepsilon_d
-V^2 g_0(\varepsilon)}.
\end{equation}
The energy and momentum widths of the spectral function at a given energy near
the Dirac point can be approximated as
$\Delta\varepsilon_\mathbf{k}\equiv{\hbar}/\tau_\mathbf{k}=-2\,
\mathrm{Im}\,\Sigma(\varepsilon_\mathbf{k})$, and
$\hbar\Delta{k}=-(2/v_F)\,\mathrm{Im}\,\Sigma(\varepsilon_\mathbf{k})$,
respectively, where $\tau_\mathbf{k}$ is the scattering time and $v_F$ is the
Fermi velocity.  Energy and momentum widths, which are proportional to $n_i$
at low concentrations according to Eq.~(\ref{ata}), can be deduced from ARPES
energy distribution curves (EDC) and momentum distribution curves (MDC),
respectively, and can be used to estimate the density of randomly distributed
H adsorbates using these results.  In Fig.~\ref{fig3}, we plot the spectral
function for pure graphene and for graphene with a relatively large 2\%
impurity concentration to make the effects large enough to be more clearly
visible. Our plots compare well with the experimental results of
Ref.~\onlinecite{haberer2010}.  We can observe the effect of spectral
broadening and the fingerprints of the localized state below the band
boundary at $-7.8$~eV.  Less trivial are the changes observed around
the Dirac point, i.e., the zero energy at $K$ point, where both sublattice
symmetry breaking and the presence of the midgap resonance can play a role.
Even within the simplest approximation that we have used to relate the
many-impurity to the single-impurity scattering, we see a quasi band-gap
opening and the appearance of new quasilocalized states within the gap near
zero energy.
\cite{pereira2006, skrypnyk2011}

In conclusion, we presented analytic expressions for the LDOS of graphene with
a single hydrogen atom adsorbate, as well as for the spectral function of
graphene with a low density of randomly distributed hydrogen adsorbates.  Our
plots of the spectral function explain the features of quasi band gap and the
spectral line broadening observed in angle-resolved photoemission spectra of
hydrogenated graphene. Our expressions can be used to estimate the
concentration of hydrogen adsorbates from measured spectral linewidths.

We had stimulating discussions with S.~A.~Jafari. M.~F.
thanks H.~Rafii-Tabar for kind support. A.~G. and D.~H. acknowledge the DFG
Grant No.~GR 3708/1-1. A.~G. acknowledges an APART fellowship from the
Austrian Academy of Sciences.

\end{document}